\documentclass[conference]{IEEEtran}
\usepackage{cite}
\ifCLASSINFOpdf
   \usepackage[pdftex]{graphicx}
\else
\fi
\usepackage{amsmath}
\usepackage{array}
  \usepackage[caption=false,font=footnotesize]{subfig}
\usepackage{multirow,makecell}
\usepackage{amsthm}
\usepackage{amssymb}
\usepackage{color}

\onecolumn

\begin{document}
\title{On Discrete Age of Information of Infinite Size Status Updating System}

\author{\IEEEauthorblockN{Jixiang Zhang, Yinfei Xu}
\IEEEauthorblockA{School of Information Science and Engineering, 
Southeast University, NanJing, China.\\
Email: \{230179077, yinfeixu\}@seu.edu.cn.}}
%\and
%\IEEEauthorblockN{Homer Simpson}
%\IEEEauthorblockA{Twentieth Century Fox\\
%\and
%\IEEEauthorblockN{James Kirk\\ and Montgomery Scott}
%\IEEEauthorblockA{Starfleet Academy\\
%San Francisco, California 96678--2391\\


%Telephone: (800) 555--1212, Fax: (888) 555--1212}


\maketitle

\begin{abstract}
In this paper, for the discrete time status updating system with infinite size, we derive the explicit expression of average age of information (AoI), and the AoI's stationary distribution. Based on the discussions of difference between finite and infinite size status updating systems, we successfully characterize the random dynamics of system's AoI using a two-dimensional state vector, which simultaneously tracks the real time AoI and the age of packet currently under service. We constitute the two-dimensional age process and completely solve all the stationary probabilities. Then, as one of the marginal distributions, the stationary distribution of system's AoI is derived, and with which the mean of the AoI is also determined.  
\end{abstract}


% \fi
\IEEEpeerreviewmaketitle

\newtheorem{Definition}{Definition}  
\newtheorem{Theorem}{Theorem}  
\newtheorem{Lemma}{Lemma}
\newtheorem{Corollary}{Corollary}
\newtheorem{Proposition}{Proposition}
\newtheorem{Remark}{Remark}

\section{Introduction}

\iffalse
This demo file is intended to serve as a ``starter file''
for IEEE conference papers produced under \LaTeX\ using
IEEEtran.cls version 1.8b and later.
I wish you the best of success.

\hfill mds
 
\hfill August 26, 2015

\subsection{Subsection Heading Here}
Subsection text here.

\subsubsection{Subsubsection Heading Here}
Subsubsection text here.
\fi

Age of information (AoI) of a status updating system was proposed in paper \cite{1}, which was defined as the time elapsed since the generation time of the last received packet in the destination. In recent years, it has been used widely as the freshness metric to characterize the timeliness of various communication systems. A detailed summary of main contributions of AoI was given in paper \cite{2}. So far, plenty of results have been obtained which determine the mean of AoI for the system with different queue models, even in some works, for example in work \cite{3}, the property of AoI's stationary distribution was considered. In papers \cite{4,5,6}, for simple queues including $M/M/1$, $M/D/1$, and $D/M/1$, the expression of system's average AoI was obtained explicitly. Packet management strategies were discussed in paper \cite{7}, and the benefit of introducing proper packet deadlines, both deterministic and random to reduce the long term average AoI was proved in \cite{8} and \cite{9}. Recently, many papers have been launched considering the AoI of simple status updating networks, such as the system with multiple sources \cite{10} and the system having multi-hop packet transmission \cite{11,12}.

On the other hand, although there are not many, but still have some works analyzing the AoI of discrete time systems. In paper \cite{13}, using the proof techniques and tools developed for analyzing continuous AoI, the authors obtained the average (peak) AoI of status updating system with $Ber/G/1$ and $G/G/\infty$ queues. Later, in work \cite{14} the expression of discrete AoI's distribution was obtained (in form of Laplace-Stieltjes Transform (LST)) for basic system with different service disciplines. Discrete time system with multiple sources was studied in paper \cite{15}, where the authors obtained the exact per-source distribution of AoI and peak AoI in matrix-geometric form. In our work \cite{16}, we obtain the explicit formula of average discrete AoI for bufferless status updating system by defining a two-dimensional age process, which characterizes the AoI and the age of packet in service as a whole.

In this paper, we consider the stationary AoI of an infinite size status updating system which uses $Ber/Geo/1/\infty$ queue, and derive the explicit expressions of AoI's mean and its distribution. Based on the characteristic of infinite size system, it shows that the random dynamics of system's AoI can be described using a two-dimensional stochastic process. We establish the stationary equations of defined age process and obtain all the stationary probabilities by solving the equations completely. Then, as one of the marginal distributions, the stationary distribution of AoI can be determined as well. Given AoI's distribution, the mean of the AoI can be calculated easily.

The rest of the paper are organized as follows. In Section \ref{sec2}, we depict the model of an infinite size status updating system and describe system's age of information. In Section \ref{sec3}, the difference of finite and infinite size system is discussed, based on which we successfully characterize the dynamics of AoI using a two-dimensional age process. Then, explicit expressions of both AoI's mean and its stationary distribution are derived. Numerical simulations including AoI's distributions and the cumulative probabilities are plotted in Section \ref{sec4}. In addition, we also compare the expressions of average discrete AoI with that of corresponding continuous AoI, and propose a possible relationship between average AoIs $\overline{\Delta}_{Ber/Geo/1/c}$ and $\overline{\Delta}_{M/M/1/c}$ for general system size $c$. At last, the paper is concluded in Section \ref{sec5}.

\section{System Model and Problem Formulation}\label{sec2}
\begin{figure}[!t]
\centering
\includegraphics[width=3.6in]{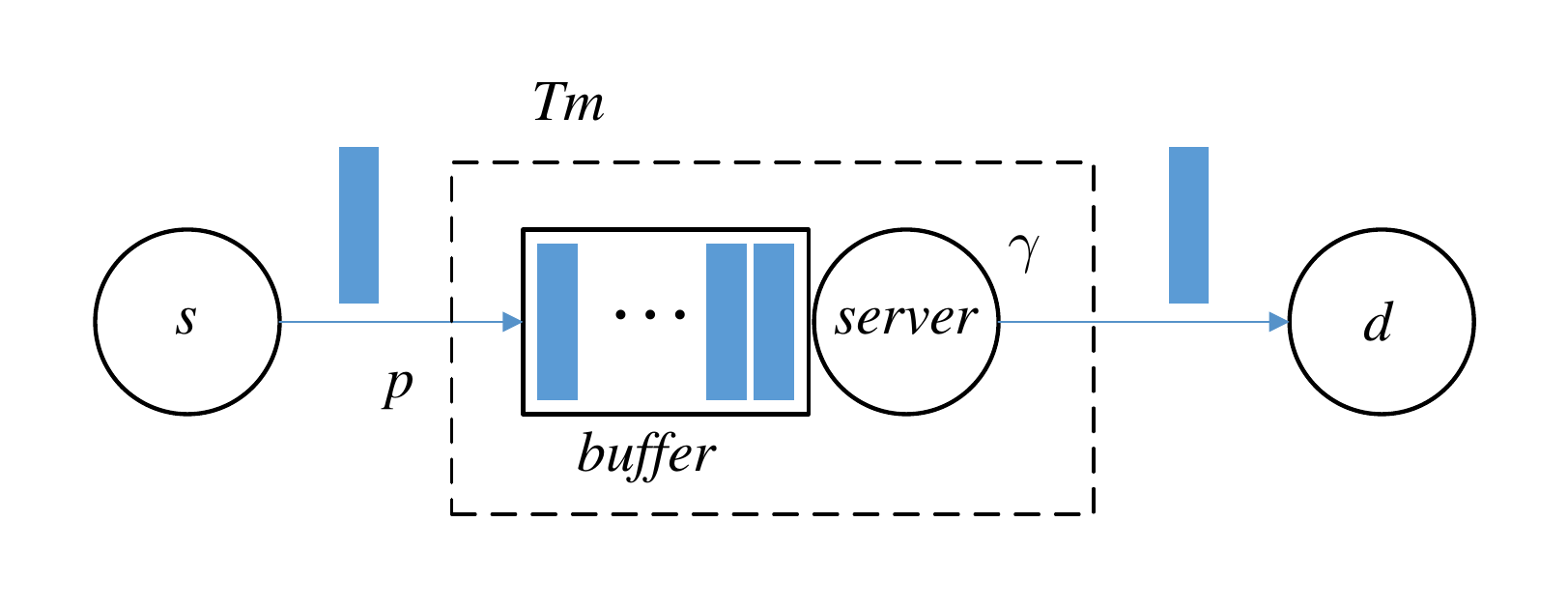}
% where an .eps filename suffix will be assumed under latex, 
% and a .pdf suffix will be assumed for pdflatex; or what has been declared
% via \DeclareGraphicsExtensions.
\caption{The model of an infinite size discrete time status updating system.}
\label{fig_sys}
\end{figure}

In Figure \ref{fig_sys}, we depict the model of an infinite size status updating system. Packet arrivals to the transmitter is assumed to form a Bernoulli stochastic process, and the packet service time follows the geometric distribution with intensity $\gamma$. Updating packet generated at $s$ is transmitted to the destination $d$ through the transmitter, in which a random period of time is consumed. The service discipline of the system is First-Come First Served (FCFS). The age of information (AoI) at the destination $d$ is defined as the time elapsed since the generation time of the last received packet up to now. During the time when no packet is obtained, the value of AoI increases 1 after each time slot ends. Every time when a packet passes the transmitter and arrives to $d$, the value of the AoI will be reduced to the \emph{system time} of the received packet, which is actually equal to the instantaneous age of this packet.

Let $a(k)$ be the value of AoI in $k$th time slot. Then, the AoI at next time slot, $a(k+1)$ is determined by 
\begin{equation}\label{E1}
a(k+1)=
\begin{cases}
a(k)+1           &  \text{ if no packet is obtained at the receiver,}   \\
a(k)+1-Y_j    &  \text{ if $j$th packet arrives to $d$.}
\end{cases}
\end{equation}
where $Y_j$ is the interarrival time between $(j-1)$th and $j$th arriving packet. Since the system considered here has infinite capacity, thus no packet is missing and these $(j-1)$th and $j$th packets are truly generated successively in the source node. However, when the size of system is finite, no matter how small the probability is, some packets are deleted when they arrive and find that the system is full.

The time average AoI is defined as follows, which converges to the mean of AoI because it is assumed that the age process is ergodic. We use the notation $\Delta_{inf}$ to represent the stationary AoI of the infinite size status updating system. Then, we have 
\begin{align}
\overline{\Delta}_{inf}&=\lim\nolimits_{T\to\infty}\frac{1}{T}\sum\nolimits_{k=1}^{T}a(k)  \notag   \\
&=\lim\nolimits_{T\to\infty}\frac{1}{T}\sum\nolimits_{i=1}^{M_T}i\cdot\left|\left\{1\leq k\leq T:a(k)=i \right\}\right|  \label{E2} \\
&=\sum\nolimits_{i=1}^{\infty}i\cdot\pi_i    \label{E3}
\end{align}
in which $M_T=\max_{1\leq k\leq T}a(k)$, and for each $i\geq 1$, 
\begin{equation}\label{E4}
\pi_i=\lim\nolimits_{T\to\infty}\frac{\left|\left\{1\leq k\leq T:a(k)=i \right\}\right|}{T}
\end{equation}
is the probability that stationary AoI takes value $i$. Notice that probability distribution $\left\{\pi_i,i\geq1\right\}$ is the stationary distribution of discrete AoI $\Delta_{inf}$, and from which AoI's mean can be calculated according to equation (\ref{E3}).

The randomness of both packet arrivals and the service time in server makes the AoI at destination change randomly. After one time slot, the AoI may increase by 1 if no packet is obtained, or drops to the instantaneous age of the obtained packet if one such packet is successfully received. Although an infinite number of packets can be contained in current system, in following Section \ref{sec3} we will prove that using a two-dimensional state vector, which records the real time AoI and the age of packet in system's server, is sufficient to fully describe the random dynamics of system's AoI. As long as all the stationary probabilities of the two-dimensional age process are obtained, the distribution of AoI is determined as one of two marginal distributions. Given AoI's distribution, then the average value of AoI can be calculated.

\section{AoI of Infinite Size Status Updating System}\label{sec3}
In the following, we talk about the difference between the systems with finite and infinite size. These discussions will give the method of characterizing the AoI of an infinite size status updating system.

Observing that when the system has infinite capacity, all the packets from source node can enter the system and be stored in buffer. For finite size status updating systems, no matter how large the buffer is, it is still possible that the coming packet is discarded since the buffer is already full. In paper [3], the authors named the system without packet discarding as \emph{lossless system}, while other systems are called \emph{lossy systems}. Packet discardings make the time interval between two consecutive packets received at the destination undetermined, since some packets may been deleted when they arrive to transmitter and find the buffer full. Therefore, in order to describe the random changes of all the packet ages (including the value of AoI) in a finite size system, we have to set an age-parameter $m_i$ for each packet. However, if no packets are deleted at all times, the age difference between two consecutive packets is exactly equal to the \emph{interarrival time}, whose probability distribution is determined by the packet arrival process. For the Bernoulli arrivals with parameter $p$, suppose the age of current packet is $m_j$, then we can determine that the next packet's age $m_{j+1}$ is $m_j-IA$ with probability $(1-p)^{m_j-m_{j+1}-1}p$, where $IA$ is the random interarrival time between two successive packets. Since the age of one packet cannot be less than zero, when $IA\geq m_j$, we let $m_{j+1}=0$. For these cases, the next packet is in fact not existent because it has not yet arrived. This observation is sufficient for us to characterize all the random dynamics of the age of information at the destination node.

Define the two-dimensional state vector $(n,m)$, $n>m\geq0$ where $n$ denotes system's AoI and $m$ represents the age of packet in service. Constituting the age process corresponding to infinite size system as 
\begin{equation}\label{E5}
AoI^{(\infty)}=\left\{(n_k,m_k):n_k>m_k\geq0,k\in\mathbb{N}\right\}     
\end{equation}
When there is no packet in system, i.e., both the unbounded buffer and the server are empty, the value of $m$ is set to 0.

We list all the random transfers of $(n,m)$ in Table \ref{table1}, where the considered random variables that trigger the state vector transitions are the interarrival time $IA$ and the service variable $B$, which represents whether the packet service is completed in one time slot. Also define $\pi_{(n,m)}$ as the stationary probability of state vector $(n,m)$ where $n>m\geq0$. We establish all the stationary equations of age process $AoI^{(\infty)}$ as follows.  
\begin{equation}\label{E6}
\begin{cases}
\pi_{(n,m)}=\pi_{(n-1,m-1)}(1-\gamma) + \sum\nolimits_{k=n}^{\infty}\pi_{(k,n-1)}p\gamma(1-p)^{n-m-1}  &  \quad  (n>m\geq2)   \\
\pi_{(n,1)}=\pi_{(n-1,0)}p(1-\gamma) + \sum\nolimits_{k=n}^{\infty}\pi_{(k,n-1)}p\gamma(1-p)^{n-2}   & \quad   (n\geq2)   \\
\pi_{(n,0)}=\pi_{(n-1,0)}(1-p) + \sum\nolimits_{k=n}^{\infty}\pi_{(k,n-1)}\gamma(1-p)^{n-1}   &  \quad   (n\geq2)   \\
\pi_{(1,0)}=\sum\nolimits_{k=1}^{\infty}\pi_{(k,0)}p\gamma   
\end{cases}
\end{equation}

\begin{table}[!t]
%% increase table row spacing, adjust to taste
\renewcommand{\arraystretch}{1.3}
% if using array.sty, it might be a good idea to tweak the value of
% \extrarowheight as needed to properly center the text within the cells
\caption{Descriptions of state vector transfers of age process $AoI^{(\infty)}$}
\label{table1}
\centering
%% Some packages, such as MDW tools, offer better commands for making tables
%% than the plain LaTeX2e tabular which is used here.
\begin{tabular}{c|c|c}
\Xhline{1pt}
Initial state     &  Considered r.v.s   &  Realizations of $IA$, $B$ and next state      \\
\Xhline{1pt}
\multirowcell{6}{$(n,m)$,\\$n>m\geq1$}    &         $B=0$          &    $(n+1,m+1)$    \\
\cline{2-3}      &      $(B=1,IA)$       &     $(IA=1):(m+1,m)$       \\
\cline{3-3}      &                            &     $(IA=2):(m+1,m-1)$       \\
\cline{3-3}      &                            &            \\
      &                            &     $\vdots$       \\
      &                            &            \\
\cline{3-3}      &                            &     $(IA=m):(m+1,1)$       \\
\cline{3-3}      &                            &     $(IA\geq m+1):(m+1,0)$       \\
\hline
\multirowcell{3}{$(n,0)$,\\$n\geq1$}          &         $IA\geq 2$         &    $(n+1,0)$     \\
\cline{2-3}      &    $(IA=1,B)$       &       $(B=0):(n+1,1)$     \\
\cline{3-3}      &                              &        $(B=1):(1,0)$     \\
\Xhline{1pt}
\end{tabular}
\end{table}

Equations (\ref{E6}) are explained as follows. First of all, we consider how the state vector $(n,m)$, $n>m\geq 2$ can be transferred to. From initial state vector $(n-1,m-1)$, if the packet service does not finish in current time slot, then apparently $(n,m)$ is obtained at next time slot. On the other hand, beginning with an arbitrary state vector of form $(k,n-1)$, $k\geq n$, as long as following two conditions are satisfied, that is, (i) the packet service is completed; and (ii) the interarrival time $IA$ between current and next packets is equal to $n-m$, then the state of age process $AoI^{(\infty)}$ will transfer to $(n,m)$ as well. Notice that when the value of AoI is reduced to $n$, $IA=n-m$ implies that the age of packet which will be served next time is equal to $m$. Combining these two cases, we obtain the first row of (\ref{E6}).

For the cases $n\geq2$, next the stationary equations of $(n,1)$ are discussed. Before one time slot, assuming the state vector is $(n-1,0)$, it is observed that a packet arrives and stays in system after one time slot will make the state vector change to $(n,1)$. In addition, from any state vector $(k,n-1)$ where $k\geq n$, when the packet service ends after one time slot, the packet of age $n$ is delivered to destination, which reduces the AoI to $n$. Meanwhile, let the interarrival time $IA$ between current and next coming packet equal $n-1$, then the age 1 packet at the head of queue will enter system's server. As a result, again we obtain the state vector $(n,1)$ at next time slot. This explains the second row of equations (\ref{E6}).

The stationary equations for state vectors $(n,0)$, $n\geq 2$ are analyzed according to similar discussions. In this situation, the system is empty and the current value of AoI is $n$. The state vector $(n-1,0)$ jumps to $(n,0)$ as long as no packet arrives to the transmitter at the beginning of next time slot. Apart from this, similarly, $(n,0)$ can also be obtained from the state vectors having form $(k,n-1)$. Let the packet in server is the unique one in system, when this packet completes the service and leaves, and no packet comes at next time slot, the entire system is emptied. This requires the interarrival time $IA$ is long enough, more specifically, $IA$ has to be larger than $n$, the transition probability is equal to 
\begin{equation}
\gamma\sum\nolimits_{k=n}^{\infty}(1-p)^{k-1}p=\gamma(1-p)^{n-1}   \notag  
\end{equation}
Thus, we see that the state vector $(k,n-1)$ transfers to $(n,0)$. Summarizing all these possibilities, the third row of equations (\ref{E6}) are determined.

Finally, a packet enters an empty system and consumes only one time slot in system's server will offer the receiver one packet of age 1. Then, the AoI is reset to 1 and notice that this is the only way the state vector $(1,0)$ can be obtained.

In the following, we determine all the stationary probabilities $\pi_{(n,m)}$, $n>m\geq0$ by solving the system of equations (\ref{E6}) directly.    
\begin{Theorem}
The stationary probabilities $\pi_{(n,m)}$, $n>m\geq 0$ of age process $AoI^{(\infty)}$ are determined as  
\begin{equation}\label{E7}
\pi_{(n,0)}=\frac{p}{1-p}\Big\{ (1-p)^n-(1-\gamma)^n  \Big\}     \qquad    (n\geq 1)   
\end{equation}
and for $n>m\geq1$, 
\begin{align}\label{E8}
\pi_{(n,m)}=\frac{p^2}{1-p}(1-p)^n \left(\frac{1-\gamma}{1-p}\right)^m - p\gamma(1-\gamma)^{n-1} + \frac{p(\gamma-p)}{(1-p)(1-\gamma)}\left(\frac{1-\gamma}{1-p}\right)^m(1-\gamma)^{n-m} 
\end{align}
\end{Theorem}

\begin{IEEEproof}
Now, we completely solve the system of equations (\ref{E6}) and derive all the stationary probabilites $\pi_{(n,m)}$, $n>m\geq0$. First of all, for $n>m\geq2$, repeatedly using the first row of (\ref{E6}) gives   
\begin{align}
\pi_{(n,m)}&=\pi_{(n-1,m-1)}(1-\gamma) + \sum\nolimits_{k=n}^{\infty}\pi_{(k,n-1)}p\gamma(1-p)^{n-m-1}  \notag \\
&=\left\{ \pi_{(n-2,m-2)}(1-\gamma) + \sum\nolimits_{k=n-1}^{\infty}\pi_{(k,n-2)}p\gamma(1-p)^{n-m-1} \right\}(1-\gamma) 
+ \sum\nolimits_{k=n}^{\infty}\pi_{(k,n-1)}p\gamma(1-p)^{n-m-1}  \notag \\
&=\pi_{(n-2,m-2)}(1-\gamma)^2 + p\gamma(1-p)^{n-m-1} \left\{ (1-\gamma) \sum\nolimits_{k=n-1}^{\infty}\pi_{(k,n-2)} +  \sum\nolimits_{k=n}^{\infty}\pi_{(k,n-1)}  \right\}   \notag  \\
& \quad \qquad \qquad \qquad \qquad \qquad \qquad \qquad \qquad \qquad \vdots   \notag \\
&=\pi_{(n-m+1,1)}(1-\gamma)^{m-1} + p\gamma(1-p)^{n-m-1} \left(\sum\nolimits_{j=0}^{m-2}(1-\gamma)^j \sum\nolimits_{k=n-j}^{\infty}\pi_{(k,n-1-j)} \right)   \notag  \\
&=\left\{\pi_{(n-m,0)}p(1-\gamma)+\sum\nolimits_{k=n-m+1}^{\infty}\pi_{(k,n-m)}p\gamma(1-p)^{n-m-1}\right\}(1-\gamma)^{m-1}    \notag \\
&\qquad \qquad \qquad  \qquad \qquad \quad + p\gamma(1-p)^{n-m-1} \left( \sum\nolimits_{j=0}^{m-2}(1-\gamma)^j \sum\nolimits_{k=n-j}^{\infty}\pi_{(k,n-1-j)} \right)     \label{E9}  \\
&=\pi_{(n-m,0)}p(1-\gamma)^m + p\gamma (1-p)^{n-m-1} \left( \sum\nolimits_{j=0}^{m-1}(1-\gamma)^j \sum\nolimits_{k=n-j}^{\infty}\pi_{(k,n-1-j)}  \right)  \label{E10}
\end{align}
where in (\ref{E9}), the boundary case given in the second row of (\ref{E6}) is substituted.

Equation (\ref{E10}) is used to determine the probabilities $\pi_{(n,m)}$, $n>m\geq1$ in the end, after we obtain the explicit formulas of $\pi_{(n,0)}$, $n\geq1$ and the infinite sums $\sum\nolimits_{n=m+1}^{\infty}\pi_{(n,m)}$ for each $m\geq1$.

Define
\begin{equation}\label{E11}
f_m=\sum\nolimits_{n=m+1}^{\infty}\pi_{(n,m)}   \qquad  (m\geq1) 
\end{equation}
we derive the exact expressions of $f_m$ in the followings.

Let $n>m\geq2$, using the relations in stationary equations (\ref{E6}), it shows that 
\begin{align}
f_m&=\sum\nolimits_{n=m+1}^{\infty}\pi_{(n,m)} \notag \\
&=\sum\nolimits_{n=m+1}^{\infty}\left\{\pi_{(n-1,m-1)}(1-\gamma)  + \sum\nolimits_{k=n}^{\infty}\pi_{(k,n-1)} p\gamma(1-p)^{n-m-1} \right\}  \notag \\
&= \left(\sum\nolimits_{n=m}^{\infty}\pi_{(n,m-1)} \right)(1-\gamma) + p\gamma  \sum\nolimits_{n=m+1}^{\infty}(1-p)^{n-m-1}\left(\sum\nolimits_{k=n}^{\infty}\pi_{(k,n-1)} \right)   \notag  \\
&=f_{m-1}(1-\gamma) + p\gamma  \sum\nolimits_{n=m+1}^{\infty}(1-p)^{n-m-1}f_{n-1}   \label{E12}
\end{align}

Do once iteration, we obtain the relation 
\begin{equation} \label{E13}
f_{m-1}=f_{m-2}(1-\gamma) + p\gamma  \sum\nolimits_{n=m}^{\infty}(1-p)^{n-m}f_{n-1} 
\end{equation}

Using equations (\ref{E12}) and (\ref{E13}), computing following difference 
\begin{align}
&(1-p)f_m-f_{m-1}  \notag  \\
={}&f_{m-1}(1-p)(1-\gamma) + p\gamma\sum\nolimits_{n=m+1}^{\infty}(1-p)^{n-m}f_{n-1} - f_{m-2}(1-\gamma) - p\gamma\sum\nolimits_{n=m}^{\infty}(1-p)^{n-m}f_{n-1}   \notag  \\
={}&f_{m-1}(1-p)(1-\gamma)  -p\gamma f_{m-1}  - f_{m-2}(1-\gamma)   \label{E14}
\end{align}
which is equivalent to  
\begin{equation}\label{E15}
(1-p)f_m - [(1-p)+(1-\gamma)]f_{m-1}+(1-\gamma)f_{m-2}=0 
\end{equation}

Notice that (\ref{E15}) is a order two difference equation with constant coefficients, its characteristic equation is 
\begin{equation}
(1-p)r^2 - [(1-p)+(1-\gamma)]r+(1-\gamma)=(r-1)[(1-p)r-(1-\gamma)]=0  
\end{equation}
and we can immediately obtain the two roots of the equation as  
\begin{equation}
r_1=1, \quad r_2=\frac{1-\gamma}{1-p}.    
\end{equation}

According to the theory of difference equations, the general expression of $f_m$ can be written as $f_m=c_1r_1^m + c_2r_2^m$, where $c_1$ and $c_2$ are the undetermined coefficients. Since all the stationary probabilities add up to 1, we have  
\begin{equation}\label{E18}
1 =\sum\nolimits_{m=0}^{\infty}\sum\nolimits_{n=m+1}^{\infty}\pi_{(n,m)} = \sum\nolimits_{m=0}^{\infty}f_m   
\end{equation}
in which $f_0$ can be defined similarly as $f_0=\sum\nolimits_{n=1}^{\infty}\pi_{(n,0)}$.

Therefore, if $|r_i|\geq 1$, the corresponding coefficient $c_i$ must be zero. Otherwise, the infinite sum (\ref{E18}) cannot converge. As a result, we conclude that 
\begin{equation} \label{E19}
f_m=c\left(\frac{1-\gamma}{1-p}\right)^m   \qquad   (m\geq 1)  
\end{equation}

To determine $f_m$ completely, we have to find the first sum $f_0$ and the unkonwn coefficient $c$ in equation (\ref{E19}). 

Let $m=1$, from the second row of (\ref{E6}), $f_1$ is calculated as  
\begin{align}
f_1&=\sum\nolimits_{n=2}^{\infty}\pi_{(n,1)} \notag \\
&= \sum\nolimits_{n=2}^{\infty}\left\{\pi_{(n-1,0)}p(1-\gamma) + \sum\nolimits_{k=n}^{\infty}\pi_{(k,n-1)} p\gamma(1-p)^{n-2} \right\} \notag \\
&=p(1-\gamma)f_0 + p\gamma\sum\nolimits_{n=2}^{\infty}(1-p)^{n-2}f_{n-1}  \notag \\
&=p(1-\gamma)f_0 + p\gamma\sum\nolimits_{n=2}^{\infty}(1-p)^{n-2}c\left(\frac{1-\gamma}{1-p}\right)^{n-1} \notag \\
&=p(1-\gamma)f_0+ cp\frac{1-\gamma}{1-p}  \label{E20}
\end{align}

Using the general expression (\ref{E19}) and combining with (\ref{E20}), we obtain that 
\begin{equation}
f_1=c\frac{1-\gamma}{1-p}=p(1-\gamma)f_0+ cp\frac{1-\gamma}{1-p}     \notag
\end{equation}
which gives $f_0=c/p$.

Finally, since the sum of all the stationary probabilities is 1, we have following equations 
\begin{align}
1&=\sum\nolimits_{m=0}^{\infty}\sum\nolimits_{n=m+1}^{\infty}\pi_{(n,m)}   \notag   \\
&=\sum\nolimits_{n=1}^{\infty}\pi_{(n,0)} + \sum\nolimits_{m=1}^{\infty}\sum\nolimits_{n=m+1}^{\infty}\pi_{(n,m)}   \notag  \\
&=f_0 + \sum\nolimits_{m=1}^{\infty}f_m  \notag  \\
&=\frac{c}{p} + \sum\nolimits_{m=1}^{\infty} c\left(\frac{1-\gamma}{1-p}\right)^m  \notag  \\
&=\frac{c}{p}+\frac{c(1-\gamma)}{\gamma-p}  
\end{align}
from which we can solve that $c=\frac{p(\gamma-p)}{(1-p)\gamma}$.

So far, all the $f_m$, $m\geq0$ are determined explicitly. We show that 
\begin{equation}\label{E22}
f_m=\sum\nolimits_{n=m+1}^{\infty}\pi_{(n,m)}=
\begin{cases}
\frac{\gamma-p}{(1-p)\gamma}   &   \quad (m=0)  \\
\frac{p(\gamma-p)}{(1-p)\gamma}\left(\frac{1-\gamma}{1-p}\right)^m    &    \quad (m\geq 1)
\end{cases}
\end{equation}

Given the results (\ref{E22}), now the probabilities $\pi_{(n,0)}$ can be determined. The third row of equations (\ref{E6}) shows    
\begin{align}
\pi_{(n,0)}&=\pi_{(n-1,0)}(1-p) + f_{n-1}\gamma(1-p)^{n-1}    \notag  \\
&=\pi_{(n-1,0)}(1-p) + \frac{p(\gamma-p)}{1-p}(1-\gamma)^{n-1}     \qquad   (n\geq2)    \label{E23}
\end{align}

Applying equation (\ref{E23}) iteratively, we have 
\begin{align}
\pi_{(n,0)}&=\left(\pi_{(n-2,0)}(1-p) + \frac{p(\gamma-p)}{1-p}(1-\gamma)^{n-2}\right)(1-p) + \frac{p(\gamma-p)}{1-p}(1-\gamma)^{n-1}    \notag  \\
&=\pi_{(n-2,0)}(1-p)^2 +\frac{p(\gamma-p)}{1-p} \Big\{ (1-p)(1-\gamma)^{n-2}-(1-\gamma)^{n-1} \Big\}  \notag \\
& \qquad \qquad \qquad \qquad \qquad \qquad  \vdots  \notag \\
&=\pi_{(1,0)}(1-p)^{n-1} + \frac{p(\gamma-p)}{1-p}\sum\nolimits_{j=0}^{n-2}(1-p)^j(1-\gamma)^{n-1-j}  \notag \\
&=p\gamma f_0 (1-p)^{n-1} + \frac{p(\gamma-p)}{1-p}\sum\nolimits_{j=0}^{n-2}(1-p)^j(1-\gamma)^{n-1-j}   \label{E24} \\
&= \frac{p(\gamma-p)}{1-p}\sum\nolimits_{j=0}^{n-1}(1-p)^j(1-\gamma)^{n-1-j}  \notag  \\
&=\frac{p}{1-p}\Big\{(1-p)^n - (1-\gamma)^n\Big\}   \label{E25}
\end{align}
where in (\ref{E24}) we have used the last equation in (\ref{E6}), which says $\pi_{(1,0)}=p\gamma f_0$. Then, expression given in (\ref{E22}) is substituted.

Since (\ref{E25}) holds for $n\geq2$, we have to check that equation (\ref{E25}) also gives $\pi_{(1,0)}$ when we let $n=1$. This verification is direct and we conclude that the probability expression (\ref{E25}) is valid for all $n\geq1$.

At last, by equation (\ref{E10}), the other probabilities $\pi_{(n,m)}$, $n>m\geq1$ can be obtained as well. We have  
\begin{align}\label{E229}
\pi_{(n,m)}=\pi_{(n-m,0)}p(1-\gamma)^m + p\gamma(1-p)^{n-m-1} \left(\sum\nolimits_{j=0}^{m-1}(1-\gamma)^jf_{n-1-j}\right) 
\end{align}

Since both the probability $\pi_{(n-m,0)}$ and $f_{n-1-j}$ are already known, substituting them into (\ref{E229}) and finally we show that the probabilities $\pi_{(n,m)}$ are determined to be  
\begin{align}
\pi_{(n,m)}=\frac{p^2}{1-p}(1-p)^{n-m}(1-\gamma)^m - p\gamma(1-\gamma)^{n-1}  + \frac{p(\gamma-p)}{(1-p)(1-\gamma)}\left(\frac{1-\gamma}{1-p}\right)^m(1-\gamma)^{n-m} \quad (n>m\geq1)
\end{align}

So far, all the stationary probabilities are solved and we complete the proof of Theorem 1. 
\end{IEEEproof}

Since the AoI is denoted by the first component of state vector, then the probability that stationary AoI equals each $n$ can be obtained by merging all the probabilities $\pi_{(n,m)}$ which have identical AoI-component $n$.  
\begin{Theorem}
Assuming the queue model in discrete time status updating system is Ber/Geo/1/$\infty$, then the stationary distribution of AoI $\Delta_{Ber/Geo/1/\infty}$ is given as 
\begin{align}
&\Pr\{\Delta_{Ber/Geo/1/\infty}=n\}   \notag  \\
={}& \frac{p\gamma}{\gamma-p}(1-p)^n-\left[\frac{p^2(1-\gamma)}{\gamma-p}+\gamma \right](1-\gamma)^{n-1} + \frac{\gamma-p}{1-p}\left(\frac{1-\gamma}{1-p}\right)^{n-1}-p\gamma(n-1)(1-\gamma)^{n-1}  \quad  (n\geq1)  \label{E28}
\end{align}
and the average AoI $\overline{\Delta}_{Ber/Geo/1/\infty}$ is calculated to be 
\begin{align}
\overline{\Delta}_{Ber/Geo/1/\infty}&=\frac{(1-p)(p+\gamma)}{p(\gamma-p)} + \frac{p(1-\gamma)(p-2\gamma)}{\gamma^2(\gamma-p)} - \frac{1}{\gamma}  \label{E29} \\
&=\frac{1}{\gamma}\left((1-\gamma) + \frac{1}{\rho_d} +\frac{\rho_d^2(1-\gamma)}{1-\rho_d} \right)  \label{E30}
\end{align}
in which $\rho_d=p/\gamma$ is the discrete traffic intensity. 
\end{Theorem}

\begin{IEEEproof}
The probability that the AoI equals $n$ can be obtained by collecting all the stationary probabilities $\pi_{(n,m)}$ where $0\leq m \leq n-1$. For $n\geq 2$, we have 
\begin{equation}
\Pr\{\Delta_{Ber/Geo/1/\infty}=n\} = \pi_{(n,0)} + \sum\nolimits_{m=1}^{n-1}\pi_{(n,m)}    \notag  
\end{equation}
where the latter sum is equal to  
\begin{align}
&\frac{p^2}{1-p}(1-p)^n \sum\nolimits_{m=1}^{n-1}\left(\frac{1-\gamma}{1-p}\right)^m - p\gamma(n-1)(1-\gamma)^{n-1} + \frac{p(\gamma-p)}{(1-p)(1-\gamma)}(1-\gamma)^n \sum\nolimits_{m=1}^{n-1}\left(\frac{1}{1-p}\right)^m    \notag  \\
={}&\frac{p^2(1-\gamma)}{\gamma-p}\Big\{(1-p)^{n-1} - (1-\gamma)^{n-1} \Big\} - p\gamma(n-1)(1-\gamma)^{n-1} - \frac{\gamma-p}{1-p}\left\{ (1-\gamma)^{n-1} - \left(\frac{1-\gamma}{1-p}\right)^{n-1} \right\}   \label{E31}
\end{align}

Combining equations (\ref{E7}) and (\ref{E31}), after some extra calculations, we show that the result (\ref{E28}) is obtained. Notice that also we have to verify that when $n=1$, expression (\ref{E28}) reduces to the probability $\pi_{(1,0)}$ because it is the probability that AoI takes value 1. This verification is direct and very easy. Therefore, we have obtained the stationary distribution of AoI explicitly for the situation where the status updating system uses $Ber/Geo/1/\infty$ queue.

Given the stationary distribution of AoI, its average value can be obtained easily by the expectation formula. We show that  
\begin{align}
\overline{\Delta}_{Ber/Geo/1/\infty}&=\sum\nolimits_{n=1}^{\infty}n\cdot \Pr\left\{\Delta_{Ber/Geo/1/\infty}=n\right\}   \notag  \\
&=\frac{p\gamma}{\gamma-p}\sum\nolimits_{n=1}^{\infty} n(1-p)^n- \left[\frac{p^2(1-\gamma)}{\gamma-p}+\gamma \right] \sum\nolimits_{n=1}^{\infty}n(1-\gamma)^{n-1}   \notag   \\
&\qquad + \frac{\gamma-p}{1-p}\sum\nolimits_{n=1}^{\infty}n\left(\frac{1-\gamma}{1-p}\right)^{n-1}-p\gamma\sum\nolimits_{n=2}^{\infty}n(n-1)(1-\gamma)^{n-1}  \label{E32}   \\
&=\frac{(1-p)(p+\gamma)}{p(\gamma-p)}+\frac{p(1-\gamma)(p-2\gamma)}{\gamma^2(\gamma-p)}-\frac{1}{\gamma}  \label{E33} \\
&=\frac{(1-\rho_d\gamma)(\rho_d\gamma+\gamma)}{\rho_d\gamma(\gamma-\rho_d\gamma)}+\frac{\rho_d\gamma(1-\gamma)(\rho_d\gamma-2\gamma)}{\gamma^2(\gamma-\rho_d\gamma)}-\frac{1}{\gamma}  \notag  \\  
&=\frac{1-\rho_d\gamma-\rho_d^2(1-\gamma)+\rho_d^3(1-\gamma)}{\gamma\rho_d(1-\rho_d)}   \notag  \\
&=\frac{\rho_d(1-\gamma)(1-\rho_d) + (1-\rho_d) + \rho_d^3(1-\gamma)}{\gamma\rho_d(1-\rho_d)}    \notag  \\
&=\frac{1}{\gamma}\left((1-\gamma) + \frac{1}{\rho_d} + \frac{\rho_d^2(1-\gamma)}{1-\rho_d}\right)    \label{E34} 
\end{align}

Equations (\ref{E33}), (\ref{E34}) gives the results (\ref{E29}) and (\ref{E30}) in the Theorem. Notice that during the computations of equation (\ref{E32}), the formula 
\begin{equation}
\sum\nolimits_{n=k}^{\infty}n(n-1)\cdots(n-k+1)(1-\xi)^{n-k}=\frac{k!}{\xi^{k+1}} \qquad  (k\geq1)
\end{equation}
is used directly to calculate two sums, where the real number $0<\xi<1$. For instance, the last sum in (\ref{E32}) is determined as 
\begin{equation}
\sum\nolimits_{n=2}^{\infty}n(n-1)(1-\gamma)^{n-1}=(1-\gamma)\sum\nolimits_{n=2}^{\infty}n(n-1)(1-\gamma)^{n-2}=\frac{2(1-\gamma)}{\gamma^3}    \notag  
\end{equation}

This completes the proof of Theorem 2. 
\end{IEEEproof}

\section{Numerical Simulations}\label{sec4}
\begin{figure*}[!t]
\centering
\subfloat[Distribution of AoI for size 1 and $\infty$ system]{\includegraphics[width=2.37in]{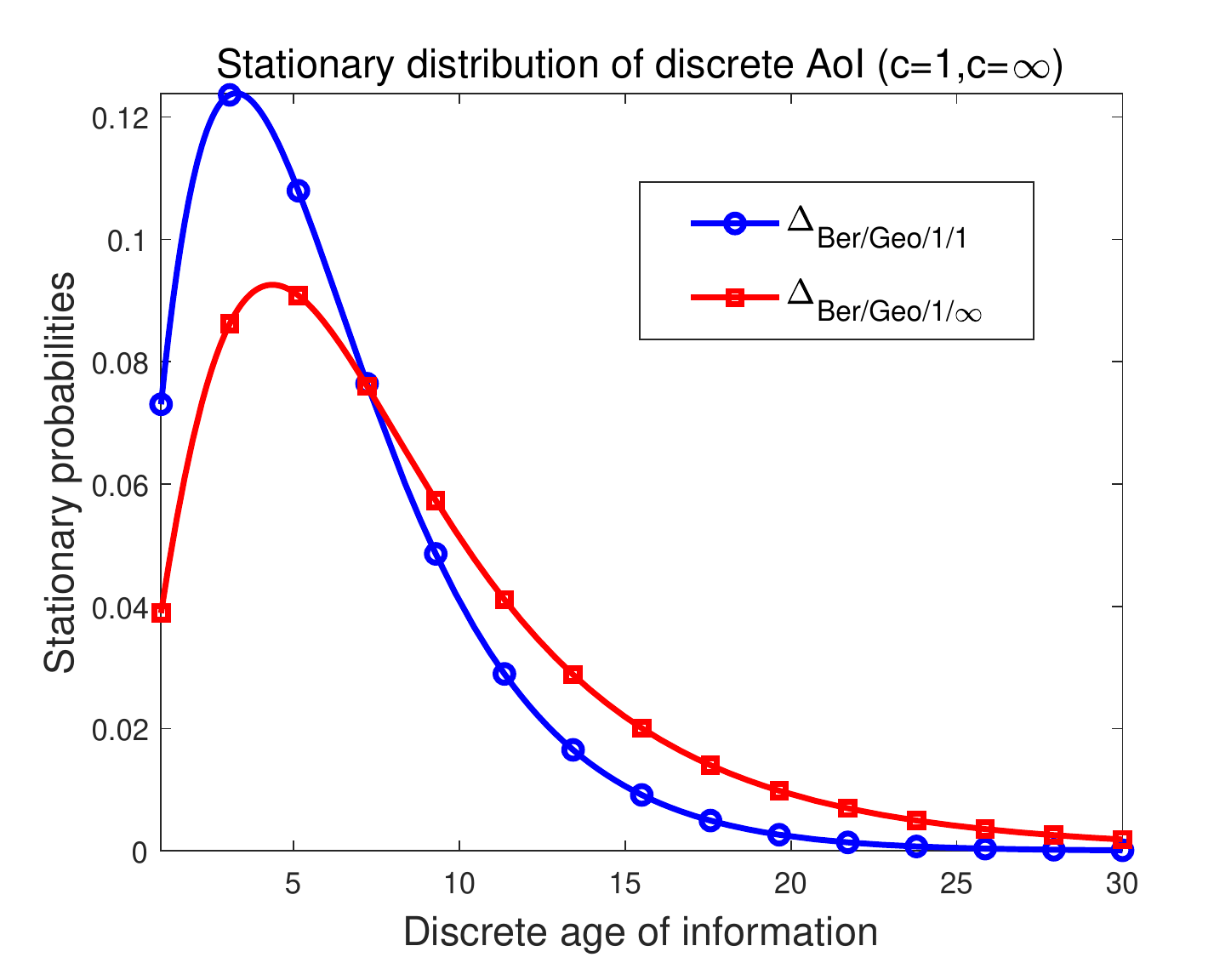}%
\label{fig_11}}
\hfil
\subfloat[Cumulative probabilities of two AoIs]{\includegraphics[width=2.37in]{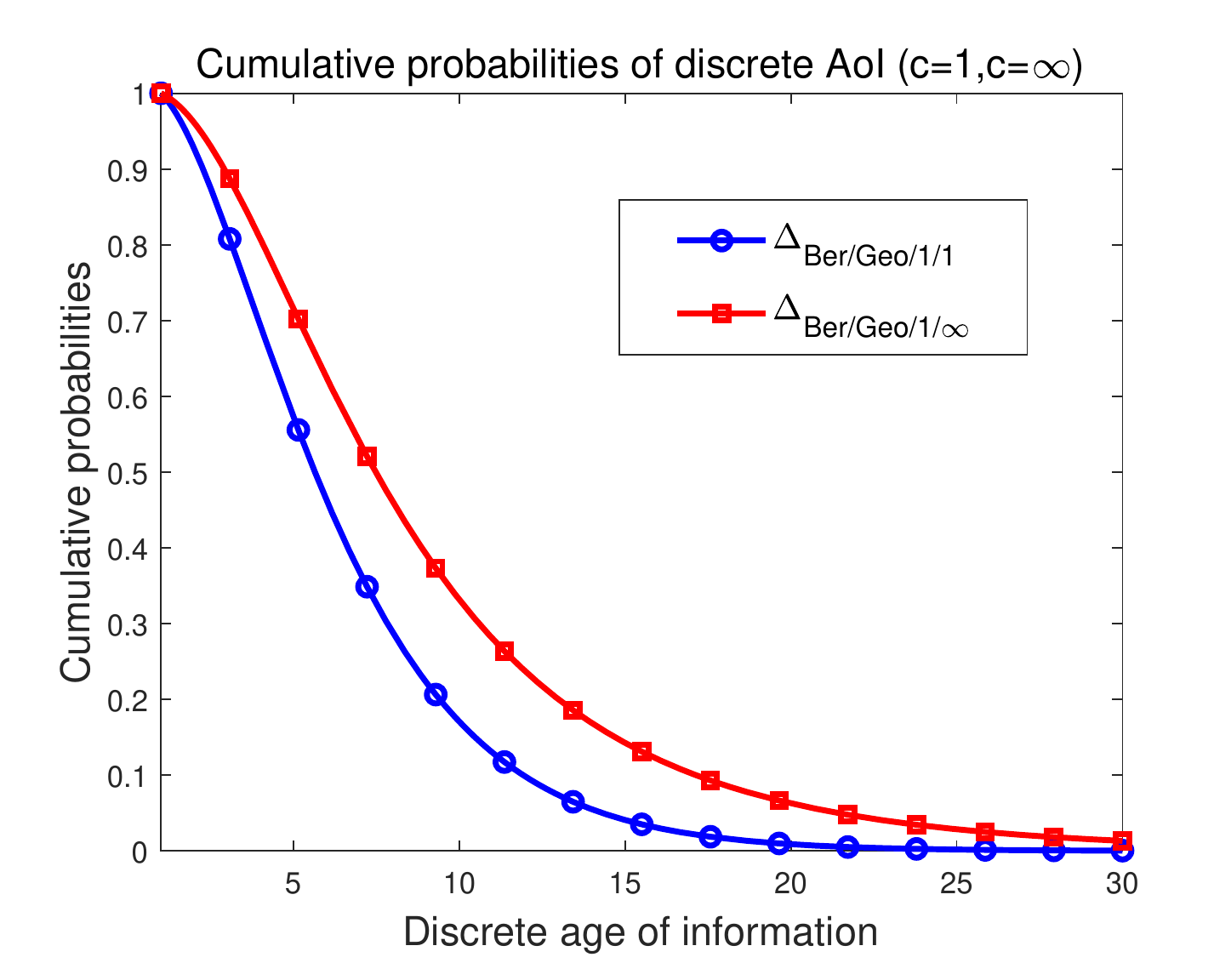}%
\label{fig_12}}
\hfil
\subfloat[Average continuous and discrete AoI]{\includegraphics[width=2.37in]{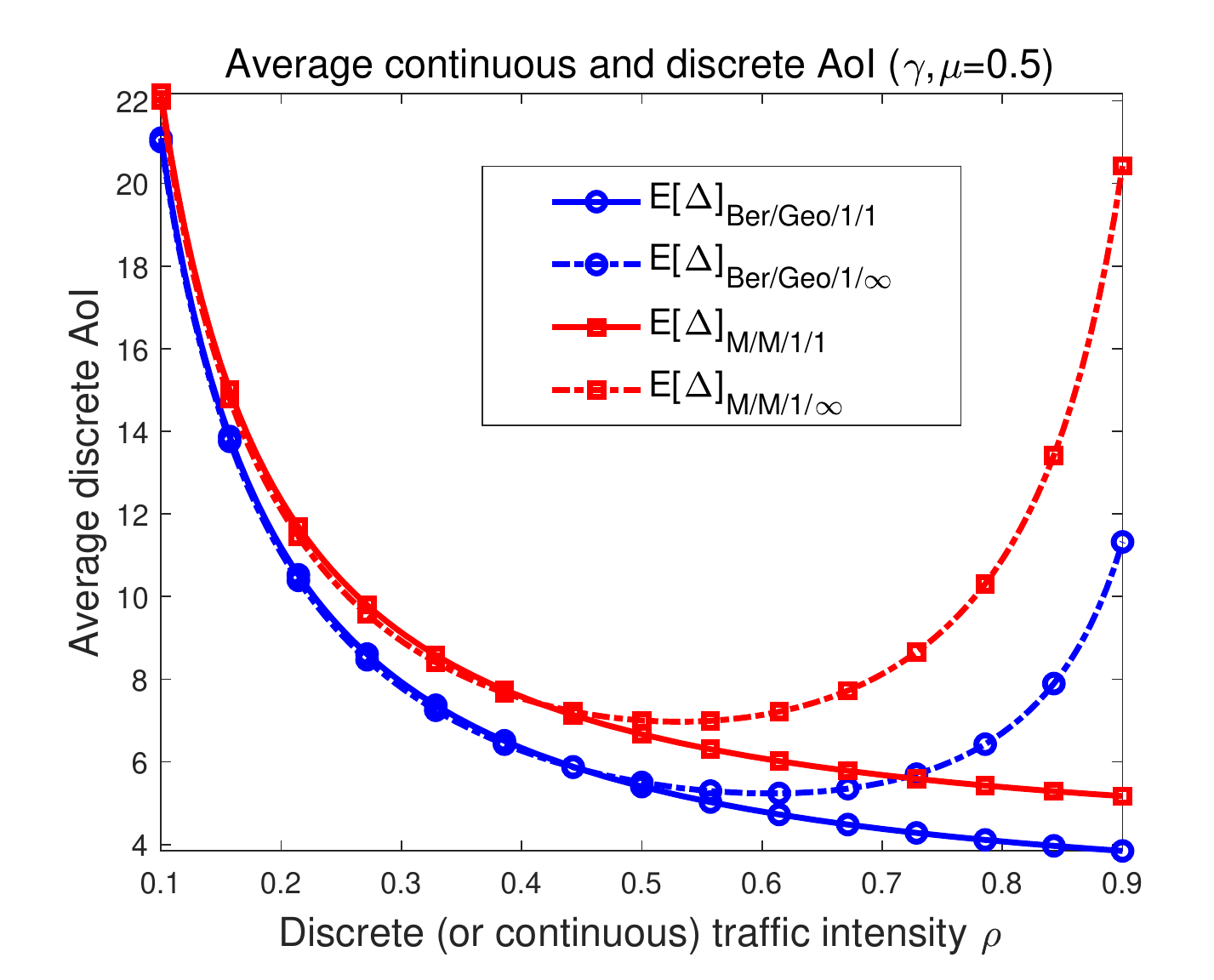}%
\label{fig_13}}
\caption{Numerical results of discrete AoI for the systems with size 1 and $\infty$ and the comparison of average continuous and discrete AoI.}
\label{fig_aoi}
\end{figure*}

In Figure \ref{fig_aoi}, we depict the graphs of AoI's stationary distribution and the corresponding cumulative probabilities for the status updating system with size 1 and size $\infty$, where the AoI's mean and its distribution for size 1 system was obtained in our work \cite{16}. We proved that 
\begin{align}
\Pr\left\{\Delta_1=n\right\}&=\frac{p(1-p)\gamma^3[(1-p)^n-(1-\gamma)^n]}{(p+\gamma-p\gamma)(\gamma-p)^2} - \frac{(p\gamma)^2n(1-\gamma)^n}{(p+\gamma-p\gamma)(\gamma-p)}   \label{E36}
\end{align}
and 
\begin{equation}\label{E37}
\overline{\Delta}_1=\frac{1}{\gamma}\left((1-\gamma)+\frac{1}{\rho_d}+\frac{\rho_d}{1/(1-\gamma)+\rho_d}\right)
\end{equation}

In addition, in paper \cite{7} the average continuous AoI of size 1 system and infinite size system were determined as   
\begin{equation}
\overline{\Delta}_{M/M/1/1}=\frac{1}{\mu}\left(1+\frac{1}{\rho}+\frac{\rho}{1+\rho}\right)   \label{E38}  
\end{equation}
and 
\begin{equation}
\overline{\Delta}_{M/M/1/\infty}=\frac{1}{\mu}\left(1+\frac{1}{\rho}+\frac{\rho^2}{1-\rho}\right)  \label{E39} 
\end{equation}

According to equations (\ref{E30}), and (\ref{E37})-(\ref{E39}), for two extreme cases where system's size is 1 and $\infty$, we plot the curves of AoI's mean versus traffic load $\rho$ and compare the average discrete and average continuous AoI. Numerical results reveal following facts: 

(i) when traffic load is small, typically lower than 0.5, the gap between continuous AoI and its corresponding discrete AoI is insignificant, while as $\rho$ gets large further, for both cases, the difference of discrete and continuous AoI becomes large; 

(ii) as $\rho$ becomes large, the mean of AoI, both continuous and discrete, are monotonically decreasing for size 1 system, however, for infinite size system, the curve of average AoI first falls down and then raises. There is an optimal $\rho^*$ (has known $\approx0.53$) such that the mean of AoI is minimized. For the system with general size $c$, there must exist the \emph{critical size} $c^*$, such that when $c<c^*$, the average AoI is always decreasing when $\rho$ becomes large, but for $c\geq c^*$, the curve has a valley, i.e., an optimal $\rho^*$ exists. It is interesting to determine this critical size $c^*$, because for given service rate $\gamma$, according to the criterion that increasing $\rho$ is always helpful for reducing average AoI, it gives a dichotomy among the status updating systems who use $Ber/Geo/1/c$ queues.

Actually, for the system with general $Ber/Geo/1/c$ queue, $c=1$ and $c=\infty$ are two extreme cases. Combining the results in \cite{16} and in current paper, we have determined the ``boundary" at both sides for AoI's mean and AoI's stationary distribution. Thus, in Figure (\ref{fig_12}), the curve of AoI's cumulative probabilities of a size $c$ system is between the red and blue lines, and in Figure (\ref{fig_13}), two blue lines give the upper and lower bounds of average discrete AoI $\Delta_{Ber/Geo/1/c}$ of system with general size $c$.

Expressions (\ref{E30}) and (\ref{E37})-(\ref{E39}) show that certain similarities exist between average continuous and average discrete AoI. Loosely speaking, we give the following possible relationship   
\begin{equation}
\mu\cdot\overline{\Delta}_{M/M/1/c} = \left.\gamma\cdot \overline{\Delta}_{Ber/Geo/1/c}\right|_{\gamma=0} \text { then replacing $\rho_d$ by $\rho$}   
\end{equation}

\section{Conclusion}\label{sec5}
In this paper, for the infinite size status updating system with $Ber/Geo/1/\infty$ queue, we determine the explicit expression of both AoI's stationary distribution and the mean of AoI. The core observation is that for a \emph{lossless} system, the age difference between successive packets is equal to the packet interarrival time. Thus, the age of next packet can be determined by the age of current packet and the interarrival time. A two-dimensional age process is constituted to describe the random changes of system's AoI, and all the stationary probabilities are obtained by solving the stationary equations directly. Combining the previous results about size 1 system and the results obtained in current paper, for status updating system with $Ber/Geo/1/c$ queues, we determine the ``boundary" at both sides for AoI's mean and its stationary distribution.

\bibliographystyle{IEEEtran}
\bibliography{paperlist.bib}
\iffalse

\fi

\end{document}